\documentclass[prb,9pt,twocolumn,superscriptaddress]{revtex4-1}
\usepackage{amsmath,amssymb,graphicx,epstopdf,braket}
\usepackage{hyperref}

\begin{document}

\newcommand{\ICFO}{ICFO -- Institut de Ciencies Fotoniques,
The Barcelona Institute of Science and Technology,
08860 Castelldefels (Barcelona), Spain}
\title{Absolute frequency references at 1529 nm and 1560 nm using modulation transfer spectroscopy}

\author{Y. Natali Martinez de Escobar}
\altaffiliation{Current address: yenny.martinez@halliburton.com}
\affiliation{\ICFO}
\author{Silvana Palacios \'{A}lvarez}
\affiliation{\ICFO}
\author{Simon Coop}
\affiliation{\ICFO}
\author{Thomas Vanderbruggen}
\affiliation{\ICFO}
\author{Krzysztof T. Kaczmarek}
\affiliation{Clarendon Laboratory, University of Oxford, Parks Road, Oxford, OX1 3PU, United Kingdom}
\author{Morgan W. Mitchell}
\email{morgan.mitchell@icfo.es}
\affiliation{\ICFO}
\affiliation{ICREA -- Instituci\'o Catalana de Recerca i Estudis
Avan\c{c}ats, 08015 Barcelona, Spain}

%

\date{\today}


\begin{abstract}
We demonstrate a double optical frequency reference (1529 nm and 1560 nm) for the telecom C-band using $^{87}$Rb modulation transfer spectroscopy.  The two reference frequencies are defined by the 5S$_{1/2} F=2 \rightarrow $ 5P$_{3/2} F'=3$ two-level and 5S$_{1/2} F=2 \rightarrow $ 5P$_{3/2} F'=3 \rightarrow $ 4D$_{5/2} F''=4$ ladder transitions.  We examine the sensitivity of the frequency stabilization to probe power and magnetic field fluctuations, calculate its frequency shift due to residual amplitude modulation, and estimate its shift due to gas collisions.  The short-term Allan deviation was estimated from the error signal slope for the two transitions.  Our scheme provides a simple and high performing system for references at these important wavelengths.  We estimate an absolute accuracy of $\sim$ 1 kHz is realistic.
\end{abstract}


\maketitle

Frequency standards near 1.55 $\mu$m are important for optical communications \cite{ITU12}, microwave photonics, remote sensing, interferometry \cite{csh06,lcc06} and fundamental metrology \cite{fel05}.  Acetylene molecular transitions lie in this region but are weak \cite{nsm05,emb05, hnp11,wwf13}.  Spectroscopy on the Rb 5S$\rightarrow$5P 780 nm  \cite{bmk98, msn07} and 5S$\rightarrow$5D 778 nm degenerate two-photon (i.e. single-frequency) \cite{csh06,csl04} atomic transitions have  been demonstrated using the second harmonic from lasers at 1560 nm and 1556 nm, respectively.  

Achieving frequency accuracies at or better than 10$^{-9}$ is possible using Doppler-free spectroscopy in various atomic and molecular species.  In fact, without much effort many labs achieve $<$ 1 MHz on an optical frequency of 3.85 $\times$ 10$^{-14}$ (the D2 line of Rb), which is already 2.6 $\times$ 10$^{-9}$.

\begin{figure}[t]
\centering
\includegraphics[clip=true,keepaspectratio=true,width= 3.25in]{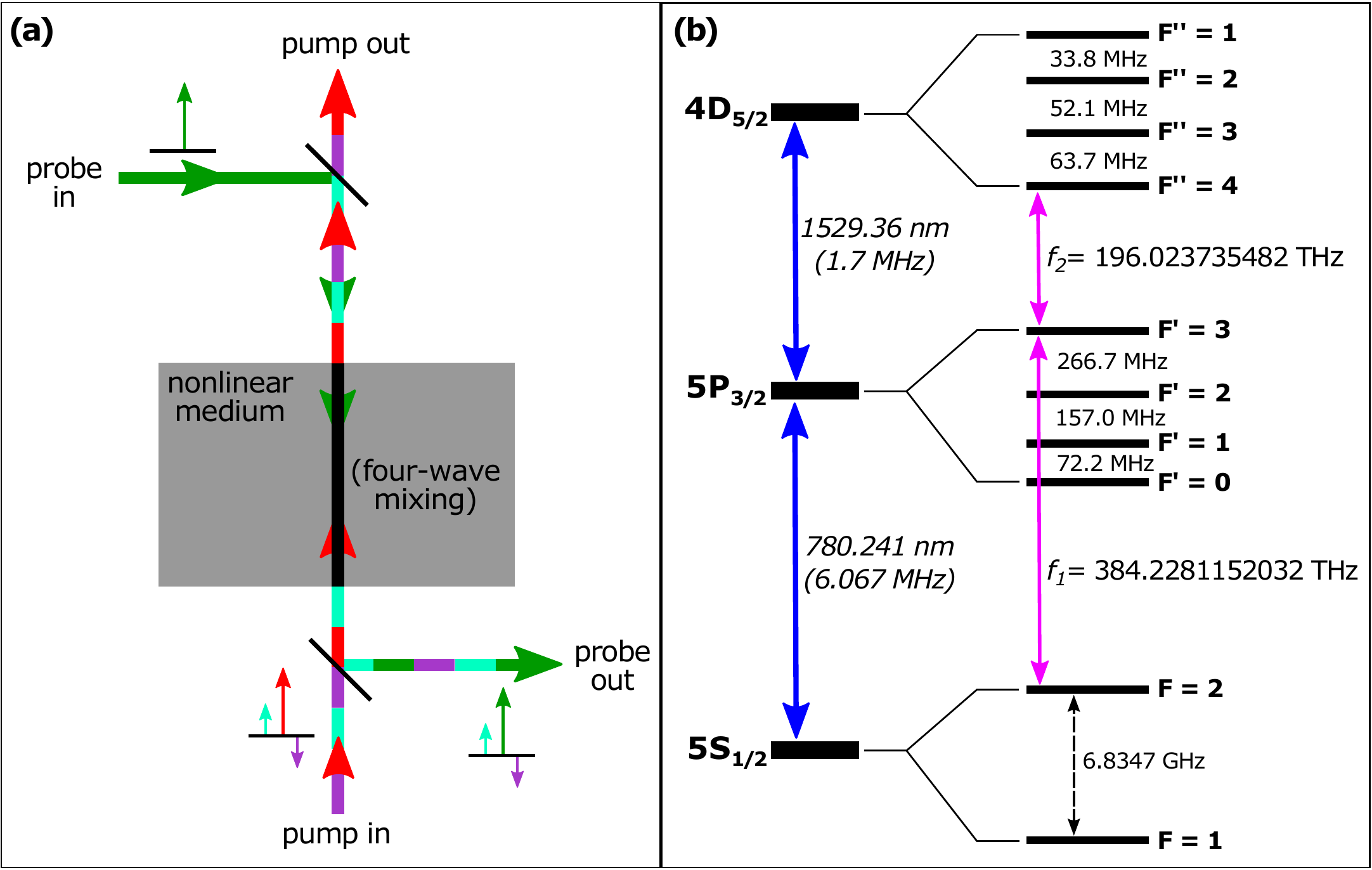}\\
\caption{(a) Representation of modulation transfer: an initially unmodulated probe beam gains modulation from a modulated pump beam via nonlinear four-wave mixing in a nonlinear medium.  (b) $^{87}$Rb energy level diagram.  Transition wavelengths and frequencies, hyperfine state splittings, and state linewidths are from \cite{lms07, nmo09,swc04, ste10}.}\label{MartinezdeEscobar_OL_fig1}
\end{figure}

Here we demonstrate absolute frequency references at 1529 nm and 1560 nm, sitting on the two edges of the telecom C-band, using modulation transfer spectroscopy (MTS) \cite{cbd82,nsm05} with $^{87}$Rb (Fig. \ref{MartinezdeEscobar_OL_fig1}).  This sub-Doppler technique offers sharp slopes, easy line identification, and modulation-free output of the stabilized lasers \cite{nsm05,mkc08,npl11}.  We demonstrate the benefits of MTS on non-degenerate two-photon transitions, and study in detail the sensitivity of the stabilized sources to residual-amplitude modulation (RAM), probe power and magnetic field fluctuations.  

In MTS, a frequency-modulated (FM) pump beam counter-propagates with a probe beam through an atomic medium as depicted in Fig. 1(a).  Resonant four-wave mixing processes transfer modulation to the initially unmodulated probe.  The resulting signals are free of linear optical background offsets.  MTS is well understood in theory \cite{cbd82,   dbl81, jaa95, shi82,  dbl82} and has been used for frequency references at 532 nm \cite{eha95, hmt99}, 612 nm \cite{gbb00}, 852 nm \cite{bcg01}, 1542 nm \cite{nsm05}, and 1560 nm \cite{mkc08, npl11}.  In the two-photon case, the line shape consists of two main components \cite{lbd86}:  one from modulation of the atomic population in the intermediate energy level (5P$_{3/2}$), and another from the two-photon coherence created between the lower and upper transitions.  Non-degenerate two-photon MTS (nMTS) was first demonstrated with Ne transitions \cite{lbd84}, and used with a 1324 nm frequency reference \cite{bct93}; the theory of nMTS is well described in \cite{lbd86}.

\begin{figure}[tbp]
\centering
\includegraphics[clip=true,keepaspectratio=true, width=0.9\columnwidth]  {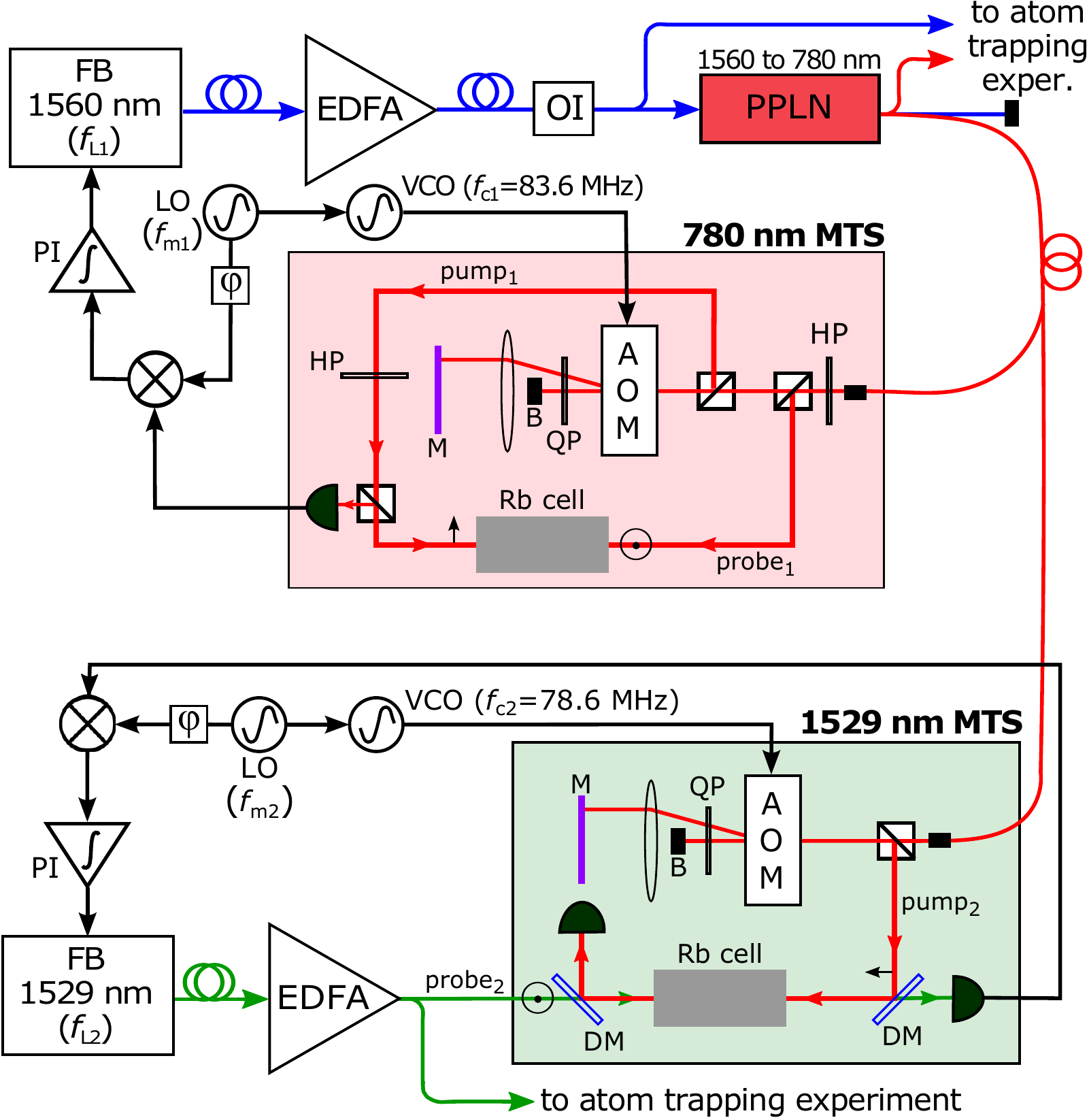}\\
\caption{Experimental setup.  FB: fiber laser, EDFA: erbium-doped fiber amplifier, OI: optical isolator, PPLN: second harmonic generation using periodically-poled LiNbO${_3}$ crystal, PI: servo lock, LO: local oscillator, VCO: voltage-controlled oscillator, AOM: acousto-optic modulator, QP: quarter-wave plate, HP: half-wave plate, DM: dichroic mirror, M: mirror, B: beam stop.  Beam polarization is indicated; the polarizations of \textsf{probe$_2$} and \textsf{pump$_2$} can be tuned with waveplates placed in their paths before entering the Rb cell (not shown).}\label{MartinezdeEscobar_OL_fig2}
\end{figure} 

The setup is illustrated in Fig. \ref{MartinezdeEscobar_OL_fig2}.  A  1560 nm fiber laser (NKT AdjustiK) with line width \textless3 kHz is amplified (NKT Koheras BoostiK) and frequency doubled to produce $\approx$ 120 mW of optical power at 780 nm.  Beams \textsf{probe$_1$}, of 0.45 mW (4.6 mW/cm$^2$), and \textsf{pump$_1$}, of 1.5 mW (15.3 mW/cm$^2$), both with 5 mm waists, are derived from this laser, the latter after double-passing an AOM.  The beams counter-propagate through a 7.2 cm long Rb glass cell (Thorlabs GC25075-RB) at room temperature without magnetic shielding.  The glass cell has a Rb pressure of > 10$^{-7}$ Torr at 25$^{\circ}$ C; its vacuum pressure before filling it with Rb was 10$^{-8}$ Torr. 

Modulating the voltage-controlled oscillator (VCO; Minicircuits ROS-80-7119) that drives the double-pass AOM with a radio-frequency (RF) local oscillator (LO) at frequency $f_{m1}$ produces the required FM for \textsf{pump$_1$}.  After the second pass in the AOM, the optical frequency of \textsf{pump$_{1}$} is shifted by twice the carrier frequency (2$f_{c1}$) and its sidebands have double the amplitude; a total optical sideband-to-carrier ratio of -10 dB requires a RF sideband-to-carrier ratio of -16 dB.  

Care was taken to reduce any measurable RF RAM caused by the VCO response envelope; this was accomplished by minimizing the power imbalance between the first-order sidebands with the spectrum analyzer for a given frequency $f_{c1}$.  As a result, $f_{c1} = 83.6$ MHz was chosen because the upper and lower sideband powers stayed equal for $f_{m1}$ ranging from 1.5 to 9 MHz.  We optimized the AOM diffraction efficiency at 83.6 MHz to center the AOM's efficiency envelope.  Note that the VCO can be replaced by a direct digital synthesizer (DDS) to improve the long-term frequency stability of LO. 

\begin{figure}[htbp]
\includegraphics[clip=true,keepaspectratio=true,width=1\columnwidth]{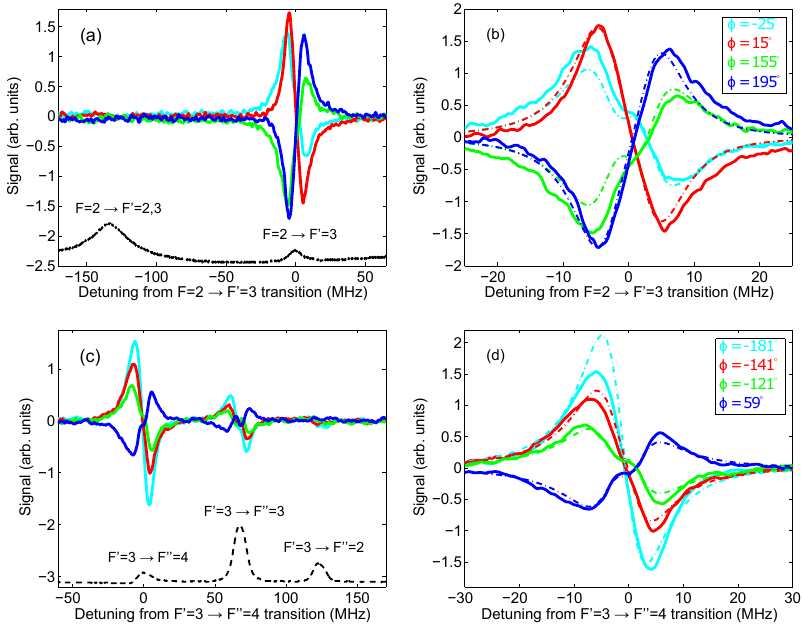}
\caption{ MTS spectra at 780 nm (upper graphs) and nMTS spectra at 1529 nm (lower graphs) in theory and experiment.  (a,c) Linear saturated absorption spectra (black dotted), plus measured lineshapes (5-pt. moving average; solid).   (b,d)  higher-resolution spectra of the strongest lines, with theoretical curves (dot-dashed).  (b) MTS spectra taken with $f_{m1}=3.413$ MHz and different demodulation phases, as given in legend.   A 3.5\% RAM was added to the FM signal to match the data.  (d) nMTS spectra taken with $f_{m2}=1.5$ MHz, $\sigma^{+}-\sigma^{+}$ polarization, and demodulation phases as given in legend.   A 0.75\% RAM was added to the FM signal to match the data.}\label{MartinezdeEscobar_OL_fig3}
\end{figure}

After exiting the cell, \textsf{probe$_1$} is detected with an amplified Si PIN photodiode and the signal demodulated with a mixer (Minicircuits ZX05-1-S).  The reference signal for demodulation at $f_{m1}$ comes from a secondary LO output, and the relative phase can be arbitrarily tuned, allowing us to explore the dispersive and absorptive properties of the atoms as a function of $f_{m1}$.  After suitable filtering and amplification we obtain sub-Doppler dispersive lineshapes (Fig. \ref{MartinezdeEscobar_OL_fig3}(a,b)) with $f_{m1}=3.413$ MHz, which we use to stabilize the frequency of the 1560 nm fiber laser.  The frequency of the locked laser can be determined by simple resonance conditions and energy conservation:  the 780 nm light is stabilized at 83.6 MHz above the $F=2 \rightarrow F'=3$ transition resonance frequency, and the 1560 nm light is stabilized at $f_{L1} = (f_{1} - f_{c1})/2$ (see Fig. \ref{MartinezdeEscobar_OL_fig1} and \ref{MartinezdeEscobar_OL_fig2} for corresponding frequencies).

For the frequency reference at 1529 nm, we used a spectroscopic setup similar to the one for 1560 nm (Fig. \ref{MartinezdeEscobar_OL_fig2}).  Our 1529 nm light comes from a second narrow-linewidth fiber laser (NKT AdjustiK) and EDFA (Keopsys CEFA-C-PB-LP B201) pair; part of this light serves as \textsf{probe$_2$}.  We demonstrate nMTS between \textsf{pump$_2$} (at 780 nm) and \textsf{probe$_2$} by counterpropagating the beams inside a second 7.2 cm long Rb cell at room temperature and no magnetic field shielding.  Dichroic mirrors allow us to combine and separate the two wavelengths (\textgreater 95\% reflectivity at 780 nm).  The powers of \textsf{pump$_2$} and \textsf{probe$_2$} before traversing the cell are 45 $\mu$W and 24 $\mu$W (4.4 mW/cm$^2$ and 3 mW/cm$^2$), respectively.

We generate and frequency modulate \textsf{pump$_2$} with a second double-pass AOM setup, and set its center frequency to reduce RAM in the same fashion as for the 1560 nm spectroscopy.  As a result, $f_{c2}= 78.6$ MHz was chosen, and equal first-order sideband power were measured with a spectrum analyzer for $f_{m2}$ ranging from 1.5 to 9 MHz.

After traversing the cell, we detect \textsf{probe$_2$} with a fast InGaAs photodiode and demodulate the signal at $f_{m2}$ with a mixer; here also the demodulation phase can be tuned.  We obtain sub-Doppler dispersive lineshapes for $f_{m2}=1.5$ MHz, which we use to stabilize the frequency of the 1529 nm laser (Fig. \ref{MartinezdeEscobar_OL_fig3}(c,d)).  By two-photon Doppler-free resonance conditions, the 1529 nm laser frequency is locked at 40.1 MHz above the $F'=3 \rightarrow F''=4$ transition ($f_{L2} = f_{2}+f_{c2}\frac{\lambda_{L1}}{\lambda_{L2}}$).  

We note three possible simplifications to the setup for compactness and portability.  First, the pumps for the MTS and nMTS could be derived from a single FM setup with a single modulation frequency.  Second, a single Rb cell could be used by laterally offsetting the MTS beams from the nMTS beams.  Finally, the AOMs could be replaced by electro-optic modulators.  

Offset-free MT spectra are shown in Fig. \ref{MartinezdeEscobar_OL_fig3} for the 1560 nm ($f_{m1}=3.413$ MHz)  and 1529 nm ($f_{m2}=1.5$ MHz) spectroscopies.  By changing the phase of the LO we alter the mixture of absorptive and dispersive contributions to the lineshape.  We use multifrequency saturation spectroscopy theory for MTS \cite{cbd82,jaa95,shi82, dbl82} and nMTS \cite{lbd86} to determine the phase corresponding to our measurements; a phase of $0^{\circ}$ corresponds to a pure in-phase contribution.  The fit required a linear combination of a pure FM signal and a small AM signal, and therefore quantified how much RAM pollutes our system \cite{jhb09}; we discuss the effects of RAM on the accuracy of the frequency reference later.  To lock the frequency of the 1560 nm laser, a more suitable MTS lineshape is found with $\phi \approx 50^{\circ}$ (modulo $180^{\circ}$), which produces maximal slope error signals \cite{jaa95,mkc08}.  For nMTS, optimal signals are found with $\phi \approx 120^{\circ}$ (modulo $180^{\circ}$).  Finally, we point out that an additional ``relaxation process'' ($\sim \Gamma_{1}/2=3$ MHz) was added to the fundamental relaxation caused by spontaneous radiative decay to account for the observed broadening of the nMTS spectra;  we attribute additional broadening mostly to transit-time broadening given the small beam  diameters ($\sim 1$ mm)  used  plus the restriction imposed by beam overlap between \textsf{pump$_2$} and \textsf{probe$_2$} for the nMTS setup.

As shown in Figs. \ref{MartinezdeEscobar_OL_fig3}(a,c), we observe strong MTS and nMTS signals only on closed transitions, as expected from theory \cite{lpn11,npl11}.  Also as expected, we found no strong nMTS lines using the $F'=2 \rightarrow F''=3 \rightarrow $ 4D$_{3/2}$ system, which contains no closed transitions.  The simplicity of the MTS and nMTS spectra simplifies interpretation and locking.

Because the atomic transition frequencies are absolute, the accuracy of the absolute frequency reference will be limited by environmental and instrumental fluctuations.  Here we study the sensitivity of the 1529 nm frequency lock to environmental magnetic fields, laser power, and the purity of the frequency modulation.  We also estimate the frequency shifts expected due to collisions.  A summary of the systematics we consider are found in Table \ref{table1}.

We first frequency stabilize the 1560 nm laser by MTS, and the 1529 nm laser by nMTS as described above; each feedback loop acts on the piezo-electric transducer (PZT) of the corresponding fiber laser, with a bandwidth of 1.2 kHz.  The feedback voltage of the 1529 nm laser, calibrated by double-resonance optical pumping (DROP) spectroscopy \cite{nmo09}, directly measures the laser's frequency displacement when operated in closed loop (PZT gain is 11.8 MHz/V in the linear regime where used).  This frequency displacement is the counteracting response of the closed loop to external perturbations, acting on any part of the full spectroscopic system, that shift the laser's frequency.  Hence, the feedback voltage indicates the full-system sensitivity to applied perturbations.  We sinusoidally modulate the magnetic field or the laser power, and de-modulate the feedback voltage to obtain the system response to perturbation.

The power modulation was applied by varying the set-point of a power-stabilization circuit controlling the 1529 nm \textsf{probe$_2$} laser power.  To modulate the ambient magnetic field, we modulated the current flowing through a solenoid wrapped around the nMTS Rb cell; monitoring the voltage across a 1 $\Omega$ resistor in series with the solenoid gave us a direct measure for the modulation amplitude applied.  Modulations were applied at 10 Hz, 100 Hz, and 500 Hz for both types of experiments.

\begin{figure}[t]
\centering
\includegraphics[clip=true,keepaspectratio=true, width=1 \columnwidth]{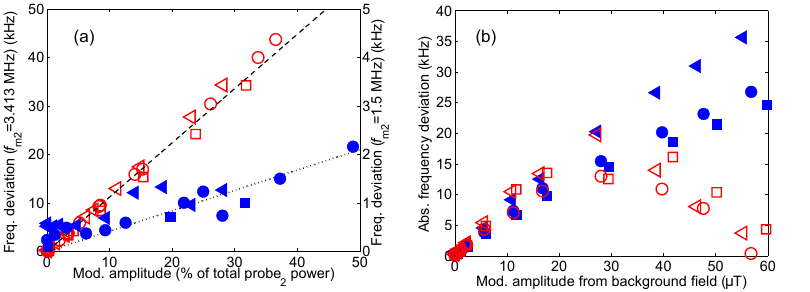} 
\caption{Sensitivity of the 1529 frequency lock to environmental perturbations. Modulation frequencies are 10 Hz (circles), 100 Hz (squares), and 500 Hz  (triangles).  (a) deviation  due to \textsf{probe$_{2}$} power modulation using $f_{m2}=1.5$ MHz (blue, solid) and 3.413 MHz (red, hollow).  Slopes:  0.042 kHz/percent power modulation  ($f_{m2}=1.5$ MHz, dotted) and 1.12 kHz/percent power modulation ($f_{m2}=3.413$ MHz, dashed).   (b) deviation due to axial magnetic field.  Using $f_{m2}=1.5$ MHz (blue, solid) and 3.413 MHz (red).}\label{MartinezdeEscobar_OL_fig4}
\end{figure}

The sensitivity of the 1529 nm frequency stabilization to power fluctuations is summarized in Fig. \ref{MartinezdeEscobar_OL_fig4}(a).  We initially lock the 1529 nm laser frequency by stabilizing \textsf{probe${_2}$} to 21 $\mu$W (2.7 mW/cm$^2$); the power of \textsf{pump${_2}$} was constant at 49 $\mu$W (4.8 mW/cm$^2$); beam polarization here is lin-perp-lin.  Figure \ref{MartinezdeEscobar_OL_fig4}(a) reveals a linear response to the imposed power modulation.  We found that an error signal using $f_{m2}=1.5$ MHz results in a more robust system for power fluctuations; theoretically, the error signal with $f_{m2}=1.5$ MHz remains linear at the zero-crossing in contrast to $f_{m2}= 3.413$ MHz.  We also point out that \textsf{pump${_2}$} powers above 100 $\mu$W (9.7 mW/cm$^2$) produced electromagnetically induced transparency (EIT)-related effects in the DROP spectra; further investigations into this are beyond the scope of this paper.

\begin{table*}[htbp]
\centering
\caption{\bf Important systematics for frequency references at 1529 nm and 1560 nm using MTS.}
\begin{tabular}{ccc}
\hline
Parameter & Max. for linewidth $<$ 1 kHz & Solution \\ [-0.025cm]
\hline
Beam power stability  & $<$ 1 \% & Noise eater \\ [-0.1cm]
Magnetic field & $<$ 1 $\mu$T & Magnetic shielding \\ [-0.1cm]
RAM & $<$ 10$^{-4}$ & cf. ref. \cite{sja13, zmb15} \\ [-0.1cm]
Background pressure & $<$ 10$^{-6}$ Torr & Vacuum environment\\ 
\hline
\end{tabular}
  \label{table1}
\end{table*}

We note that with linearly-polarized \textsf{probe${_2}$} and \textsf{pump${_2}$} beams locked on-resonance to their respective transitions, the atomic polarizability is even under time-reversal symmetry \cite{lbd86}, in contrast to the magnetic field, which is odd. We thus expect the 1529 nm frequency lock to be independent of external magnetic fields, although the transition will broaden.
Figure \ref{MartinezdeEscobar_OL_fig4}(b) shows the 1529 nm frequency lock's response to a modulated magnetic field; a significant modulation field is applied, but we only measure a small frequency shift correction, supporting our expectations.  We also performed these measurements on different days and found the frequency deviations to be similar in magnitude to that of Fig. \ref{MartinezdeEscobar_OL_fig4}(b).  Measurements of the magnetic field sensitivity imply systematic drifts from other factors such as the light polarization stability, which can differ from day.

RAM can significantly affect our error signal.  We estimate the sensitivity of both the 1529 nm and 1560 nm frequency references to RAM by calculating the theoretical frequency offset experienced by the error signal from resonance for a given amount of RAM contribution.  From our analysis we find the frequency offset experienced by the 1560 nm reference is linear (slope of linear fit is 300 kHz/\% RAM contribution using $f_{m1}=3.413$ MHz) up to a RAM contribution of 6\% of the pure FM signal;  the 1529 nm reference shift is still linear even with a 12\% RAM contribution (slope of linear fit is 81 kHz/\% RAM contribution using $f_{m2}=1.5$ MHz).  We point out that the theoretical analysis of the 1529 nm  reference excludes the additional ``relaxation" terms used previously when fitting the data (Fig. \ref{MartinezdeEscobar_OL_fig3}).

Expected frequency shifts due to collisions between Rb atoms and background gas in the cell can be estimated from previous shift measurements (Rb-Rb collisions \cite{swe80} and Rb collisions with noble gases \cite{zhr11, wni82}). At room temperature the pressure of Rb inside the cell is about 1 $\times$ 10$^{-6}$ Torr; background gas pressure is also taken to be at this level given that it is unknown.  At these pressures, the expected frequency shifts  for both Rb-Rb and Rb-noble gas collisions is $<$ 1 kHz.  A higher background pressure (e.g., He in air, about 5 ppmv or 4 mTorr) would still only cause a shift of about 2 kHz \cite{zhr11}.

The short-term Allan deviation $\sigma$($\tau$) is estimated from the peak error signal slope for both the 1560 nm and 1529 nm references \cite{hnp11}.  With our measurement bandwidth of $\sim$2 kHz, we obtain $\sigma$($\tau$) = 7.0 $\times$ 10$^{-12} \sqrt{\tau/s}$ for the 1560 nm reference, and  $\sigma$($\tau$) = 3.3 $\times$ 10$^{-12} \sqrt{\tau/s}$ for the 1529 nm reference.

We have presented an experimental and theoretical study of MTS-based telecom C-band frequency references.  In light of the measured and predicted environmental sensitivities, a 1 kHz absolute accuracy requires about 1\% power stability (with $f_{m2} = 3.413$ MHz) and $\le 1$ $\mu$T field stability, both of which are readily achieved in the laboratory by employing a noise eater and magnetic shielding.  Collision-induced shifts are controllable to under 1 kHz by employing vacuum techniques.  RAM is perhaps more critical, and must be controlled at the $10^{-4}$ level to achieve 1 kHz frequency accuracy.  Electro-optic phase modulators have shown RAM below $10^{-5}$ \cite{sja13, zmb15}, suggesting this is also achievable.

\textbf{Funding.} Spanish MINECO projects MAGO (Ref. FIS2011-23520) and EPEC (FIS2014-62181-EXP), European Research Council project AQUMET, Horizon 2020 FET Proactive project QUIC and  Fundaci\'{o} Privada CELLEX.

\end{document}